\documentstyle[twoside,fleqn]{article}

\begin{document}


\makeatletter

\topmargin -8mm
\oddsidemargin -6mm
\evensidemargin -11mm
\textheight 240mm
\textwidth 174mm
\columnsep 8mm
\columnseprule 0.2pt
\emergencystretch=6pt
\mathsurround=1pt
\mathindent=1em
\pagestyle{myheadings}
\newcommand{\bls}[1]{\renewcommand{\baselinestretch}{#1}}
\def\onecol{\onecolumn \mathindent=2em}
\def\twocol{\twocolumn \mathindent=1em}
\def\noi{\noindent}


\renewcommand{\section}{\@startsection{section}{1}{0pt}%
        {-3.5ex plus -1ex minus -.2ex}{2.3ex plus .2ex}%
        {\large\bf\protect\raggedright}}
\renewcommand{\thesection}{\arabic{section}.}
\renewcommand{\subsection}{\@startsection{subsection}{2}{0pt}%
        {-3ex plus -1ex minus -.2ex}{1.4ex plus .2ex}%
        {\normalsize\bf\protect\raggedright}}
\renewcommand{\thesubsection}{\arabic{section}.\arabic{subsection}.}
\renewcommand{\thesubsubsection}%
        {\arabic{section}.\arabic{subsection}.\arabic{subsubsection}.}

\renewcommand{\@oddhead}{\raisebox{0pt}[\headheight][0pt]{%
   \vbox{\hbox to\textwidth{\rightmark \hfil \rm \thepage \strut}\hrule}}}
\renewcommand{\@evenhead}{\raisebox{0pt}[\headheight][0pt]{%
   \vbox{\hbox to\textwidth{\thepage \hfil \leftmark \strut}\hrule}}}
\newcommand{\jheads}[2]{\markboth{\protect\small\it #1}{\protect\small\it #2}}
\newcommand{\Acknow}[1]{\subsection*{Acknowledgement} #1}


\newcommand{\Arthead}[3]{ \setcounter{page}{#2}\thispagestyle{empty}\noi
    \unitlength=1pt \begin{picture}(500,40)
        \put(0,58){\shortstack[l]{\small\it Gravitation \& Cosmology,
                        \small\rm Vol. 3 (1997), No. #1, pp. #2--#3\\
        \footnotesize \copyright \ 1997 \ Russian Gravitational Society} }
    \end{picture}          }
\newcommand{\Title}[1]{\noi {\Large #1} \\}
\newcommand{\Author}[2]{\noi{\large\bf #1}\\[2ex]\noindent{\it #2}\\}
\newcommand{\Authors}[4]{\noi
        {\large\bf #1\dag\ #2\ddag}\medskip\begin{description}
        \item[\dag]{\it #3} \item[\ddag]{\it #4}\end{description}}
\newcommand{\Rec}[1]{\noi {\it Received #1} \\}
\newcommand{\Recfin}[1]{\noi {\it Received in final form #1} \\}
\newcommand{\Abstract}[1]{\vskip 2mm \begin{center}
        \parbox{16.4cm}{\small\noi #1} \end{center}\medskip}
\newcommand{\RAbstract}[3]{ {\bf\noi #1}\\ {\bf\noi #2}
        \begin{center}\parbox{16.4cm}{\small\noi #3} \end{center}\bigskip}
\newcommand{\PACS}[1]{\begin{center}{\small PACS: #1}\end{center}}
\newcommand{\foom}[1]{\protect\footnotemark[#1]}
\newcommand{\foox}[2]{\footnotetext[#1]{#2}}
\newcommand{\email}[2]{\footnotetext[#1]{e-mail: #2}}
\def\Novg{Talk presented at the 9th Russian Gravitational Conference,
        Novgorod, 24--30 June 1996.}


\newcommand{\Ref}[1]{Ref.\,\cite{#1}}
\newcommand{\Refs}[1]{Refs.\,\cite{#1}}
\newcommand{\sect}[1]{Sec.\,#1}
\def\ten#1{\mbox{$\cdot 10^{#1}$}}
\def\npb{\nopagebreak}
\def\nq{\hspace{-1em}}
\def\nqq{\hspace{-2em}}
\def\nhq{\hspace{-0.5em}}
\def\nhh{\hspace{-0.3em}}
\def\cm{\hspace{1cm}}
\def\inch{\hspace{1in}}
\newcommand{\Figure}[2]{\begin{figure}
        \framebox[83mm]{\rule{0cm}{#1}}
        \caption{\protect\small #2}\medskip\hrule\end{figure}}


\newcommand{\sequ}[1]{\setcounter{equation}{#1}}
\def\eqdef{\stackrel{\rm def}=}
\newcommand{\Eq}[1]{Eq.\,(\ref{#1})}
\def\eq{Eq.\,}
\def\eqs{Eqs.\,}
\def\beq{\begin{equation}}
\def\eeq{\end{equation}}
\def\bear{\begin{eqnarray}}
\def\al{&\nhq}
\def\lal{&&\nqq {}}               
\def\bearr{\begin{eqnarray} \lal}
\def\ear{\end{eqnarray}}
\def\earn{\nonumber \end{eqnarray}}
\def\dst{\displaystyle}
\def\tst{\textstyle}
\newcommand{\fracd}[2]{{\dst\frac{#1}{#2}}}
\newcommand{\fract}[2]{{\tst\frac{#1}{#2}}}
\def\nn{\nonumber\\ {}}
\def\nnv{\nonumber\\[5pt] {}}
\def\nnn{\nonumber\\ \lal }
\def\nnnv{\nonumber\\[5pt] \lal }
\def\yy{\\[5pt]}
\def\yyy{\\[5pt] \lal}
\def\eql{\al =\al}

\def\e{{\,\rm e}}
\def\eps{\varepsilon}
\def\d{\partial}
\def\re{\mathop{\rm Re}\nolimits}
\def\im{\mathop{\rm Im}\nolimits}
\def\arg{\mathop{\rm arg}\nolimits}
\def\tr{\mathop{\rm tr}\nolimits}
\def\sign{\mathop{\rm sign}\nolimits}
\def\diag{\mathop{\rm diag}\nolimits}
\def\dim{\mathop{\rm dim}\nolimits}
\def\const{{\rm const}}
\def\Half{{\dst\frac{1}{2}}}
\def\half{{\tst\frac{1}{2}}}
\def\then{\ \Rightarrow\ }
\def\chg{\ \leftrightarrow\ }
\newcommand{\aver}[1]{\langle \, #1 \, \rangle \mathstrut}
\def\DAL{\raisebox{-1.6pt}{\large $\Box$}}
\newcommand{\vars}[1]{\left\{\begin{array}{ll}#1\end{array}\right.}
\newcommand{\lims}[1]{\mathop{#1}\limits}
\newcommand{\limr}[2]{\raisebox{#1}{${\lims{#2}}$}}
\def\suml{\sum\limits}
\def\intl{\int\limits}
\def\wider{\vphantom{\int}}
\def\wideup{\vphantom{\intl^a}}

\makeatother


\def\int{{\rm int}}
\def\ext{{\rm ext}}

\jheads{M.Yu. Konstantinov}
       {Causality Properties of Topologically Nontrivial
       Space-Time Models}

\bls{0.98}

\twocolumn[
\Arthead{4 (12)}{299}{304}

\Title{CAUSALITY PROPERTIES OF TOPOLOGICALLY NONTRIVIAL \yy
       SPACE-TIME MODELS\foom 1}

\Author{M.Yu. Konstantinov\foom 2}
       {Centre for Gravitation and Fundamental Metrology,
       VNIIMS, 3-1 M. Ulyanovoy Str., Moscow, 117313, Russia}

\Rec{10 September 1997}

\Abstract
{Some problems of the space-time causal structure are discussed using
models with traversable wormholes. For this purpose the conditions of
traversable wormhole matching with the exterior space-time are considered in
detail and a mixed boundary problem for the Einstein equations is
formulated and analyzed. The influence of these matching conditions
on the space-time properties and causal structure is analyzed. These
conditions have a non-dynamical nature and cannot be determined by any
physical process. So, the causality violation cannot be a result of
dynamical evolution of some initial hypersurface. It is also shown that the
same conditions which determine the wormhole joining with the outer space
provide the self-consistency of solutions and the absence of paradoxes
in the case of causality violation.}

]
\foox 1 {Talk presented at the International School-Seminar ``Problems
of Theoretical Cosmology", Ulyanovsk, September 1--7, 1997}
\email 1 {konst@rgs.phys.msu.su}


\section{Introduction}

Topologically nontrivial space-time models are described by finite or
countable sets of maps, which are joined with each other in some order \cite
{hawking73,kobayashi}. The maps joining conditions induce constraints
for the field variables. The meaning of these constraints is twofold: they
are part of the definition of geometrical objects on the manifold and must
be considered as additional boundary conditions for the field equations.

In the present paper the influence of these boundary conditions on the
causal properties of space-time will be considered for models with
traversable wormholes. For this purpose the topological structure of
space-time models with traversable wormholes will be considered and the
mixed boundary problem for Einstein equation will be formulated. Then the
applicability of the relativity principle in the exterior space-time will be
discussed for the particular case of wormhole joining. After that some
estimations for causality violation will be obtained and a simple example of
a spherical wormhole will be considered. A discussion of the ``paradoxes''
of a time travel and the so-called ``self-consistency conditions'' 
concludes the paper.

\section{The manifold structure for traversable wormholes}

The simplest space-time model with a traversable wormhole consists of two
parts: the interior space-time with the topology $M_{\int}^4=T_{\int}\times
M_{\int}^3$, and the exterior space-time $M_{\ext}^4=T_{\ext}\times
M_{\ext}^3$, where $T_{\int}$ and $T_{\ext}$ are the interior and exterior
timelike axes, $M_{\int}^3$ and $M_{\ext}^3$ are the interior and
exterior spaces, and the interior space of the wormhole $M_{\int}^3$ has
a topological structure of the direct product $M_{\int}^3=I\times M^2$ of
the interval $I=(-L_1,L_2)$ and a compact orientable 2-dimensional manifold
$M^2$ and is often called a wormhole handle. The sum $L=L_1+L_2$ is
called the coordinate length of the wormhole handle. The boundary of the
wormhole handle $M_{\int}^3$ is a disjoint sum of two manifolds $M_l^2$ and
$M_r^2$, which are often called the ``left'' and ``right'' mouths of the
wormhole and are homeomorphic in the simplest models. In the general case
$M_l^2$ and $M_r^2$ may be arbitrary compact 2-manifolds and
$M_{\int}^3$ is an interpolating manifold.

Let $t\in T_{\ext}$ and $\tau \in T_{\int}$ be the exterior and interior time
coordinates. Then, for fixed $\tau $ the wormhole connects the points of
the external spacelike hypersurface $(t_1(\tau ),M_{\ext 1}^3)$ with those
of the external spacelike hypersurface $(t_2(\tau ),M_{\ext 2}^3)$, where
$ t_1\neq t_2$ in the general case.

To discuss the causal structure of wormhole-type models, consider the
simplest case $M_l^2=$ $M_r^2=M^2$. Let $\{\tau ,\xi ^1,\xi ^2,\xi ^3\}$ be
the local coordinates in the wormhole interior, such that $-\infty <\tau
<\infty $ is an interior time-like coordinate, $\xi ^1$ is the coordinate
along the line $l$ which connects the left and right mouths of the wormhole,
$-(\sigma _1+L_1)<\xi ^1<L_2+\sigma _2$, where $\sigma _1$, $\sigma _2$, $
L_1 $ and $L_2$ are some positive constants, the values $\xi ^1=-L_1$ and
$\xi ^1=L_2 $ correspond to the left and right mouths of the wormhole,
respectively, the regions $-(\sigma _1+L_1)<\xi ^1<L_1$ and $L<\xi
^1<L+\sigma _2$ correspond to the wormhole intersection with the exterior
space-time, $\xi ^2$ and $\xi ^3$ are the coordinates on the wormhole
mouths.

In a general form the interior metric of a travers\-able wormhole may be
written in the coordinates $\{\tau ,\xi^1,$ $\xi^2,\xi^3\}$ as
\bearr
          \label{intmetric}
ds_{\int}^2=a^2(\tau ,\xi )d\tau ^2-2b_i(\tau ,\xi )d\tau
d\xi ^i-\widetilde{\gamma }_{ij}(\tau ,\xi )d\xi ^id\xi _j \nnn
\ear
where $\xi =\{\xi ^1,\xi ^2,\xi ^3\}$, $\widetilde{\gamma }_{ij}(\tau ,\xi )$
denotes the metric of the interior 3-space $\tau =\const$, $a^2(\tau ,\xi
)>0$ because of the assumption of the wormhole traversability and, in
general, $b_i(\tau ,\xi )\neq 0$.

Without loss of generality it may be supposed that in the exterior
space-time both wormhole mouths are covered by the same map $
\{t,x^1,x^2,x^3\}$, where $-\infty <t<\infty $ is an exterior time and $
\{x^1,x^2,x^3\}$ are the coordinates on the exterior space section
$t=\const$. For simplicity it will be supposed, in addition, that the
exterior space-time coordinates are synchronous, so that the exterior
space-time metric has the form
\bearr
\label{extmetr}
     ds_{\ext}^2=dt^2-\gamma _{ij}dx^idx^j
\ear
where $\gamma_{ij}$ is a metric of the exterior 3-space $t=\const$.

Let us suppose, in addition, that the exterior coordinates are comoving 
the left mouth of the wormhole, so that for $-(\sigma _1+L_1)<\xi ^1<-L_1$ we
have
\bearr                        \label{extleft}
	t_{\rm left}=\tau ,\quad x_{\rm left}^i=x^i(\xi ^1,\xi ^2,\xi ^3)
\ear
while for the right mouth, i.e. for $L_2<\xi ^1<L_2+\sigma _2$, we have in
general
\bearr
\label{extright}
	t_{\rm right}=t_r(\tau ),\quad x_{\rm right}^i=x^i(\tau ,\xi ^1,\xi
^2,\xi ^3).
\ear
These equations determine matching of the interior space-time of the
wormhole with the exterior space-time.

\section{Mixed boundary problem for wormhole models}

Eqs.\,(\ref{extleft})-(\ref{extright}) of wormhole joining with the
outer space, which determine the manifold structure, induce the following
boundary conditions for the interior and exterior metrics:
\bearr                                 \label{asympt1}
\nq  a^2(\tau ,\xi )\nnn   \nq = \left\{
     \begin{array}{lcl}
     1 & \mbox{for} & -(\sigma _1+L_1)<\xi ^1<-L_1, \yy
     \alpha ^2\cdot (1-v_iv^i) & \mbox{for} & L_2<\xi ^1<L_2+\sigma _2;
     \end{array}
               \right.\nnn \\ \lal
\label{asympt2}\nq \beta _i=\left\{
     \begin{array}{lcl}
     0 & \mbox{for} & -(\sigma _1+L_1)<\xi ^1<-L_1, \\
     \gamma _{kl}\fracd{\d x^k}{\d \tau }\fracd{\d x^l}{\d
     \xi ^i} & \mbox{for} & L_2<\xi ^1<L_2+\sigma _2;
\end{array} \right.       \nnn
\ear
and
\bearr         \label{asympt3}
\nq \widetilde{\gamma }_{ij}=\left\{
\begin{array}{ccl}
\gamma _{kl}\fracd{\d x^k}{\d \xi ^i}\fracd{\d x^l}{\d
\xi ^j} & \mbox{for} & \! -(\sigma _1+L_1)<\xi ^1<-L_1, \\
\gamma _{kl}\fracd{\d x^k}{\d \xi ^i}\fracd{\d x^l}{\d
\xi ^j} & \mbox{for}  & \! L_2<\xi ^1<L_2+\sigma _2;
\end{array}
\right.
\ear
where $x^k=x^k($ $\xi ^i)$ near the left mouth ($\xi ^1\rightarrow -(\sigma
_1+L_1)$) and $x^k=x^k($ $\tau ,\xi ^i)$ near the right mouth ($\xi
^1\rightarrow L_2$), and
\bearr
\label{boundright}\alpha =\frac{dt_r}{d\tau} ,\ \ v_l=\frac 1\alpha \gamma 
_{kl} \frac{\d x^k}{\d \tau },\ \ \beta _i=\gamma _{kl}
\frac{\d x^k}{\d \tau}\frac{\d x^l}{\d \xi ^i}.  
\ear 

The traversability condition gives the following restrictions on the
functions $t_r(\tau)$ and $x^i(\tau ,\xi )$:
\bearr
\alpha ^2(1-v_lv^l)>0,
\earn
and hence
\bearr
\label{trans}\alpha ^2>\varepsilon >0,\qquad 0\leq v_lv^l<1-\varepsilon _1
\ear
where $\varepsilon ,\varepsilon _1=\const>0$, i.e. $t_r(\tau )$ must be a
monotonic function of $\tau $ without stationary points.

\eqs (\ref{asympt1})-(\ref{asympt3}) must be considered as additional
bo\-undary conditions for the components of the metric tensor. Indeed, in 
the regions $-(\sigma _1+L_1)<\xi ^1<-L_1$ and $L_2<\xi ^1<L_2+\sigma _2$, 
where the interior space-time intersects with the exterior one, \eqs (\ref 
{extleft})-(\ref{extright}) and the induced equations 
(\ref{asympt1})-(\ref {asympt3}) have the form of coordinate transformations
and have no effect on the energy-momentum tensor. Therefore these equations
are independent of the field equations and have a nondynamical nature.

To complete the formulation of the mixed bound\-ary problem for traversable
wormhole models it is necessary to fix the field equations and initial
conditions for internal and external spaces.

For simplicity, assume that the space-time metric must
satisfy the standard Einstein equations
\bearr
\label{E00}
	G_0^0=\kappa T_0^0,\qquad G_i^0=\kappa T_i^0
\ear
and
\bearr     \label{Eij}
	G_j^i=\kappa T_j^i
\ear
where $G_\beta ^\alpha $ is the Einstein tensor, $\kappa $ is the Einstein
gravitational constant and $T_\beta ^\alpha $ is the energy-momentum tensor
of matter and non-gravitational fields. \eqs (\ref{E00}) are the
constraint equations and \eqs (\ref{Eij}) are dynamical. These
equations must be supplemented by matter and non-gravitational field
equations, not to be considered here.

The usual initial conditions for the Einstein equations must be also
specified for both interior and exterior space-times, namely: the
components of the interior metric tensor and their first partial derivatives
with respect to the interior time coordinate $\tau $, i.e.

\bearr
        \label{initint}
a^2(\tau _0,\xi ),\;\beta _i(\tau _0,\xi ),\;\widetilde{%
\gamma }_{ij}(\tau _0,\xi ),\;\d a^2(\tau _0,\xi )/\d \tau, \nnn
\beta _{i,\tau }(\tau _0,\xi ),\;\widetilde{\gamma }_{ij,\tau }(\tau_0,\xi ),
\ear
with
\bearr
a^2(\tau _0,\xi )>0,
\earn
and the components of the exterior metric and their first partial
derivatives with respect to the exterior time coordinate $t$, i.e.
\bearr
\label{initext}\gamma _{ij}(t_0,x),\quad \gamma _{ij,t}(t_0,x).
\ear
These quantities must satisfy the standard constraint equations (\ref{E00}),
the boundary conditions (\ref{asympt1})-(\ref{asympt3}) near the left mouth
($-(\sigma _1+L_1)<\xi ^1<-L_1$) and the corresponding conditions for
derivatives, which take the form
\bearr
\label{inleft1}\frac{\d a^2(\tau _0,\xi )}{\d \tau }=0,\cm
\frac{\d \beta _i(\tau _0,\xi )}{\d \tau }=0
\ear
and
\bearr
\label{inleft2}\widetilde{\gamma }_{ij,\tau }(\tau _0,\xi )=\gamma
_{kl,t}(t_0,x)x_i^kx_j^l.
\ear

So the mixed boundary problem for the wormhole models may be formulated as
follows:

\proclaim Statement.
Any space-time model with a travers\-able wormhole, whose interior and
exterior metrics have the forms (\ref{intmetric}) and (\ref{extmetr})
respectively, is a subject of the mixed boundary problem for the Einstein
equations (\ref{E00})-(\ref{Eij}), which is formed by (i) the manifold
structure equations (\ref{extleft})-(\ref{extright}), (ii) the boundary
conditions (\ref{asympt1})-(\ref{asympt3}) with the traversability
constraints (\ref{trans}), (iii) the interior and exterior initial
conditions (\ref{initint})-(\ref{initext}) which must satisfy
the boundary conditions (\ref{asympt1})-(\ref{asympt3}) and
(\ref{inleft1})-(\ref {inleft2}) near the left mouth.

An analogous mixed boundary problem may be formulated for other types of
space-time models with nontrivial topology, in particular, for space-time
models with a cosmic string which were used in \cite{gott} for time machine
construction.

\section{The relativity principle and the twin paradox for a traversable
wormhole}

In this section the particular case of wormhole joining with the exterior
space-time (\ref{extleft}) - (\ref{extright}) will be considered. Namely, it
will be assumed that $t_{\rm left}=t_{\rm right}=\tau $, i.e. $t=\tau $ is 
a global time coordinate and both mouths are placed on the same external 
space-like hypersurface. Let the wormhole mouths be also placed 
along the $z=x^1 $ axis. For simplicity we assume, in addition, 
that the external space-time is flat.

It is easy to see that, unlike the Minkowski space-time, in the 
model under consideration the exterior time $t$ will be the global time 
coordinate only in the restricted class of inertial reference frames of 
the outer space-time. As a result, the relativity principle cannot be 
applied to the motion of the wormhole mouths in the outer space. To see 
that, compare the observer motion in the outer space with respect to the
wormhole mouths with the mouths motion with respect to the observer.

In the first case the outer and inner synchronizations of events coincide in
the outer space frame of reference where the wormhole mouths are at rest 
and the observer moves. In the frame comoving the observer 
this coincidence is violated.

In the second case the interior and exterior synchronizations coincide in
the outer space frame of reference where the observer is at rest. In the 
frame comoving with one of the wormhole mouths this coincidence is 
violated.

The above considerations may be easily generalized to accelerated 
motion of the wormhole mouths, e. g. to the ``twin paradox'' 
motion.  According to the conclusion of \cite{morris,idn}, ``in the 
wormhole case the twin paradox is a true paradox involving causality 
violation''. This conclusion is based on the additional implicit 
supposition that in the interior wormhole metric the proper times of the 
mouths coincide with each other.  However, it is not necessary because the 
junction conditions (\ref{extleft}) - (\ref{extright}) are independent. In 
particular, we may consider the case when both mouths move as 
described above in the $t-z$ plane of the outer space and $t_L=t_R=\tau $. 
In this case $\tau $ is a global time coordinate in the whole space-time 
that defines the absolute synchronization of events near the wormhole 
mouths. For this reason the time delay of the right mouth relative to the 
left one is absolute and independent of the space path along which the 
comparison of clock readings is realized.

To show that it is indeed the case, recall that the time delay of the right
mouth relative to the left one is determined from the equations of the 
world lines of the comoving observers
\beq                                       \label{twin}
ds_{L}^{2}=d\tau^{2}, \quad ds_{R}^{2}=(1-V_{R}^{2})d\tau^{2}
\eeq
where $V_R=dz_R/d\tau $ is the velocity of the right mouth in the outer 
space. \eqs (\ref{twin}) have the same form for both the inner 
and outer spaces and are a special case of the equality 
(\ref{asympt1}) determining the asymptotic form for the component 
$g_{00}$ of the interior wormhole metric.  Hence the proper gravitational 
field of the right mouth induces the same time delay as the right mouth 
motion in the outer space.

We have considered only an accelerated motion of the right mouth along the
straight line. More general motion may be considered in a
similar manner. Hence we can make the general conclusion that, contrary 
to the statements of the papers \cite{morris,idn,frolnov}, accelerated 
motion of the mouths of a wormhole does not lead to its transformation 
into a time machine and to closed time-like curve (CTC) creation. So, the 
statements about unavoidable or ``absurdly easy'' wormhole transformation 
into a time machine \cite{morris,idn,frolnov,visser} are wrong. This 
conclusion conforms with the well-known theorems about the 
space-time causal structure and the Cauchy problem 
\cite{hawking73,york,fisher}.

\section{Causality violation in traversable wormhole models}

The above equations make it possible to obtain some estimates for 
causality violation in traversable wormhole models. Without loss of
generality one may assume that the mouth sizes are much smaller than
the distance between them in the outer space (the approximation of thin
mouths) and the $x^1$ axis connects the centres of the wormhole mouths.
Let, moreover, $t_r(\tau _1)>\tau _1$ for some $\tau _1>\tau _0$. Consider
a light signal sent at the moment $t_r(\tau _1)$ from the right
mouth to the left one through the wormhole and then returning to the right
mouth through the outer space. It is clear that the time delay between the
moments of signal sending and receiving is equal to
\[
     \Delta t=\delta t_1+\delta t_2-\delta t_3
\]
where $\delta t_1$ and $\delta t_2$ are the signal passing times 
through the wormhole and the exterior space, respectively, and $\delta 
t_3=t(\tau _1)-\tau _1$. The causality violation appears if 
$\Delta t\leq 0$. An estimate of the times $\delta t_1$ and $\delta 
t_2$ may be obtained from \eqs (\ref{intmetric}) and 
(\ref{extmetr}). Namely, let
\bearr
a_0^2=\min _{-L_1\leq \xi ^1\leq L_2}a^2(\tau ,\xi ),\quad
b=\max _{-L_1\leq\xi ^1\leq L_2}\left| b_i(\tau ,\xi )\right| ,\nnn
N=\max _{-L_1\leq \xi^1\leq L_2}\widetilde{\gamma }_{ij}(\tau ,\xi ),
\earn
and
\bearr
R=x^1(\tau _1,L_2),
         \quad C_{\ext}=\max _{0\leq x^1\leq R}\gamma_{11}(t,x), \nnn
C_{\int}=\frac{b+\sqrt{b^2+N}}{a_0^2},
\earn
then
\bearr
\delta t_1\leq C_{\int}L,\qquad \delta t_2\leq C_{\ext}R,
\earn
where $L=L_1+L_2$, and thus
\bearr
     \Delta t\leq C_{\int}L+C_{\ext}R-\left| t_r(\tau _1)-\tau _1\right| .
\earn
Thus a sufficient condition for causality violation may be written in
the form
\bearr                 \label{cv}
     \left| t(\tau _1)-\tau _1\right| \geq C_{\int}L+C_{\ext}R.
\ear
It is necessary to note that this estimate is very rough, so the 
causality violation may occur even if the inequality (\ref{cv}) is not 
satisfied.

Analogously, if
\bearr                               \label{causal}
     \left| t_r(\tau _1)-\tau _1\right| <C_{1\int}L+C_{1\ext}R,
\ear
where
\bearr
\nq C_{1\int}=\frac{b_m+\sqrt{b_m^2+N_m}}{a_1^2},\ \
C_{1\ext}=\min _{0\leq x^1\leq R}\gamma _{11}(t,x),
\earn
and
\bearr
a_1^2=\max _{-L_1\leq \xi ^1\leq L_2}a^2(\tau ,\xi ),\ \
b_m=\min_{-L_1\leq \xi ^1\leq L_2}\left| b_i(\tau ,\xi )\right| ,\nnn
N_m=\min_{-L_1\leq \xi ^1\leq L_2}\widetilde{\gamma }_{ij}(\tau ,\xi )
\earn
for all $\tau \in (-\infty ,\infty )$, then there are no CTCs in the
model considered.

Similar estimates may be also obtained for a wormhole with finite
mouth sizes.

Thus the main parameters which determine the causal structure of 
wormhole-type models with a given function $t_r(\tau )$ are the interior
``coordinate length'' $L$ of the wormhole handle, the exterior
``coordinate distance'' $R$ between its mouths and the factors $C_{\int}$ 
and $C_{\ext}$. It follows from the above consideration that 
the parameters $L$ and $R$ are subject to boundary conditions for 
space-time models with traversable wormhole. These parameters are 
independent of each other, of the field equations and of the function 
$t_r(\tau )$. So, using the appropriate choice of the parameters $L$ and 
$R$ (the conditions (\ref{extleft}) and (\ref{extright})) both causal and 
non-causal space-time models with traversable wormholes may be obtained 
for the same $t_r(\tau )\neq \tau $. Of course, for given boundary 
conditions (\ref{extleft})-(\ref{extright}) with $t_r(\tau )\neq \tau $ 
the causality violation depends on the factors $C_{\int}$ and $C_{\ext}$ 
determined by the field equations. On the other hand, if $t_r(\tau 
)\equiv \tau $, then a causality violation is impossible in the 
model under consideration, contrary to the statement of \cite {frolnov} 
about an unavoidable wormhole transformation into a time machine.

This confirms our earlier statements about a non-dynami\-cal nature of 
CTCs and the impossibility of dynamical wormhole transformation into a 
time machine \cite{myuk92,myuk95}.

\subsection{A spherical wormhole in Minkowski space-time}

To demonstrate that the causality violation is not directly related to
any physical processes, consider the special case of a traversable
spherical wormhole with immovable mouths, joint to flat
Minkowskian exterior space-time.

The exterior flat region of such model is described by the Cartesian 
coordinates $\left\{ t,x,y,z\right\} $, which vary from $-\infty $ to 
$\infty $, and the metric
\bearr
     ds^2=dt^2-dx^1-dy^2-dz^2,
\earn
while the interior region is described by the coordinates $\left\{\tau,
l,\theta ,\phi \right\} $, $-\infty <\tau <\infty $, $-(\sigma
_1+L_1)<l<L_2+\sigma _2$, and $(\theta ,\phi )$ are polar coordinates on
the 2-sphere $S^2$. If the wormhole mouths are placed on the 
$x$ axis with the centres at $x_{\rm left}=0$ and $x_{\rm 
right}=R=\const$, then the matching conditions 
(\ref{extleft})-(\ref{extright}) read
\bearr
     t_{\rm left}=\tau ,\,x_l=l\cos (\phi )\cos (\theta ),\nnn
          y_l=l\cos (\phi )\sin(\theta ),\,z_l=l\sin (\phi )
\earn
and
\bearr
t_{\rm right}=t_r(\tau ),\,x_r=R+l\cos (\phi )\cos (\theta ),\nnn
       y_l=l\cos (\phi)\sin (\theta ),\,z_l=l\sin (\phi ).
\earn
So the simplest interior metric which satisfies the boundary conditions 
(\ref{extleft})-(\ref{extright}) has the form
\bearr                                       \label{sphworm}
ds^2=a^2(\tau ,l)d\tau ^2-dl^2-r^2(l)(d\theta ^2+
          \sin^2(\theta )d\phi ^2), \nnn
\ear
with $a(\tau ,l)=1$ for $l<L_1$ and $a(\tau ,l)=dt_r/d\tau $ for $l>L_2$ and
$r(l)=l$ for $l<-L_1$ or $l>L_2$. In the special case when $a=a(l)$ this
metric coincides with the static metric considered in 
\cite{morris,idn,thorne}.

A direct calculation gives the following values of non-zero components of 
the Einstein tensor in the wormhole interior:
\bearr 
G_0^0=-\frac{2rr^{\prime \prime }+r^{\prime 2}-1}{r^2}, \nnn
G_1^1=-\frac{ar^{\prime 2}+2ra^{\prime }r^{\prime }-a}{ar^2}, \nnn
G_2^2=G_3^3=-\left( \frac{r^{\prime \prime }}r+\frac{a^{\prime }}a\frac{%
r^{\prime }}r+\frac{a^{\prime \prime }}a\right)
\earn
where a prime ($^{\prime }$) denotes $\d/\d l$.

It is easy to see that the Einstein tensor $G_\beta ^\alpha $ and hence the
energy-momentum tensor for the interior space in this model in 
the non-static case ($t_{\rm left}=\tau $, $t_{\rm right}=t_r(\tau )\neq 
\tau $) have the same structure and properties as in the static case 
($a(\tau ,l)=a(l)$, $ t_{\rm left}=t_{\rm right}=\tau $) which were 
considered in~\cite{thorne}. In particular, the matter in the wormhole 
interior must have the same ``exotic'' properties as in the static case. 
Further, the Einstein equations do not restrict the dependence $a(\tau 
,l)$ on the interior time $\tau $. Taking into consideration that 
$t_r(\tau )$ is an arbitrary monotonic function ($dt_r(\tau )/d\tau $ 
$\neq 0$), one may conclude that the Einstein equations (and hence the 
physical processes in the space-time) have no effect on the causal 
structure of the model.

Evidently the same result may be obtained for any model with
an arbitrary static exterior space-time, in particular, for the so-called
ringhole model, which was considered recently in \cite{diaz}.

\section{The ``paradoxes'' of time machine and ``self-consistency
conditions''}

Causality violation is associated traditionally with different paradoxes
which are usually formulated in the following way \cite{krasnikov}: 
somebody, after passing through a time machine, kills his parents, which
makes impossible his time travel. It is clear, that the ``paradox'' 
appears because the observer is considered as an object which moves along 
its world line, while in the presence of closed time-like curves such 
a consideration is incorrect and the whole world line which represents 
the object must be considered.

As an example, consider the motion of a self-interac\-ting test particle of
mass $m$ in a background with a wormhole ``time machine''. The exterior
space-time is supposed to be Minkowskian and the sizes of the wormhole
mouths are negligibly small (the point-like mouths approximation) and at 
rest in some reference frame. For definiteness, we shall assume that the 
wormhole mouths in the exterior space-time have the coordinates 
$(t,\vec r_A)$ and $(t+a,\vec r_B)$, where $a=a(t)>0$, and $a(t)\geq |\vec 
r_B-\vec r_A|$ for some time interval $t_1\leq t\leq t_2$, where $t_1$ and 
$t_2$ are some constants. So, the region $t_1\leq t\leq t_2+a(t_2)$ of the 
exterior space-time contains the paths of the closed time-like or null 
curves which violate causality (we use geometric units where $c=1$).

As in~\cite{carlini}, the following particle motion will be considered.
The particle starts at a time $t_i$ in the position $\vec r_i$, enters the
mouth (B) of the wormhole at a time $\overline{t}+a(\overline{t})$ 
(the position $ \vec r_B$), where $\overline{t}>t_1$, goes out of the 
other mouth (A) at an earlier time $\overline{t}$ (position $\vec r_A$) 
and finally ends its trajectory at a time $t_f$ in the position $\vec 
r_f$.  The path length of the wormhole handle is assumed to be infinitely 
short, so the motion through the wormhole in the proper time of the 
particle is almost simultaneous.  According to an external observer, 
the particle traversing the time machine travels back in time by the 
amount $\Delta t=-a(\overline{t})$ where $a(\overline{t})\geq |\vec 
r_B-\vec r_A|$ by assumption. For simplicity, the motion with the only 
self-intersection of the particle world line at the point with coordinates 
$(t_0,\vec r_0)$ will be considered here.

Consider the region of the exterior Minkowskian space-time with 
$\overline{t} < t < \overline{t} + a(\overline{t})$. The world line of the 
particle may be considered in this region as two world lines of two copies 
of the same particle with the positions $\vec r_1(t)$ and $\vec r_2(t)$. 
Both particles may be considered as independent objects which interact by 
means of a potential $V$ of special type. The geometry of the model 
imposes some additional limitations on the possible motion. Namely, the 
entrance of the particle into mouth B and its exit from A may be written 
formally as (\eq (3) of~\cite{carlini})
\bearr
\label{ent}
\vec r_1(\overline{t}+a(\overline{t}))=\vec r_B \\ \lal
\label{exit}\vec r_2(\overline{t})=\vec r_A
\ear
and the self-intersection of the particle world line has the form (\eq(13)
of~\cite{carlini})
\bearr
\label{intersect}
     \vec r_1(t_0)=\vec r_2(t_0)=\vec r_0.
\ear

Carlini et al. \cite{carlini} obtained an exact solution of the 
equations of motion, which correspond to a special case of the
potential $V$, with the constraints (\ref{ent})-( \ref{intersect}). It was 
stated that the existence of such a solution, which minimizes the action 
functional, shows that the ``Principle of self-consistency'' is a 
consequence of the ``Principle of minimal action''~\cite{carlini}.

Let us analyze this statement in more detail. It is clear that \eqs
(\ref{ent})-(\ref{intersect}) are a direct consequence of the definition of
the self-intersected line on the manifold. To be applied to the particle 
world line, the condition (\ref{ent}) states that the particle falls  
into the wormhole at some time $\overline{t}+a(\overline{t})$ 
independently of the previous history, and the condition (\ref{exit}) 
states that the world line $(\overline{t},\vec r_2(\overline{t}))$ is 
a continuation of the world line of the same particle. Therefore, 
according to the constraints (\ref{ent})-(\ref{intersect}), the points 
($t,\vec r_1(t))$ and $(t,\vec r_2(t))$ are points of different paths 
of the world line of the same particle. For the same reason the point 
$(t_0,\vec r_1(t_0))=(t_0,\vec r_2(t_0))=(t_0,\vec r_0) $ is a 
self-intersection point of the same world line.  Moreover, the
conditions (\ref{ent})-(\ref{intersect}) also state that the 
self-intersection of the particle world line does not prevent its 
passing through the time machine. Hence, the conditions 
(\ref{ent})-(\ref{intersect}) prevent the appearance of paradoxes which 
are usually associated with the existence of the time machine.

So, the self-consistency of the solution (the absence of some 
``paradoxes'') is provided not by the ``Principle of minimal action'', but 
by the constraint equations (\ref{ent})-(\ref{intersect}) which are 
part of the definition of the self-intersected line in the manifold.

Of course, the constraints (\ref{ent})-(\ref{intersect}) are not 
independent of the equations of motion. Namely, the number of 
the particle entrances into the time machine, the existence and number of 
self-intersections of its world line, as well as the precise values of the 
parameters $\overline{t}$, $ t_0$ and $\vec r_0$ for every 
self-intersection are determined by the equation of motion. But if the 
particle world line has a self-intersection, a local solution of 
the equations of motion near each self-intersection point must satisfy 
the geometrical constraints (\ref {ent})-(\ref{intersect}) or their 
generalization.

The above conditions (\ref{ent})-(\ref{intersect}) are pure geometrical and
are not directly related to the action functional. In the quantum case 
there appears some additional restriction on the particle world line. 
Indeed, the probability amplitude $\psi =\exp \{iS/\hbar \}$ must be a 
function of point, while the action $S$ is a function of the world line. 
Therefore, for any closed world line the action functional of a test 
particle must satisfy the additional constraint
\bearr                              \label{quant}
S=\oint Ld\sigma =0 \ {\rm mod}\ (2\pi \hbar )
\ear
where $\sigma $ is the canonical parameter on the world line. 
\eq(\ref{quant}) may be called the ``world line quantization condition''. 
This condition formally coincides with the well-known Bohr-Sommerfeld 
quantization condition but is principally different in its nature.

\Acknow
{This work was supported in part by the Russian Ministry of Science and the
Russian Fund of Basic Research (grant N 95-02-05785-a).}


\small
\begin{thebibliography}{99}

\bibitem{hawking73}  S.W. Hawking and G.F.R. Ellis, ``The Large-Scale
Structure of Space-Time'', Cambridge U. P., 1973.

\bibitem{kobayashi}  S. Kobayashi and K. Nomizu, ``Foundation of 
Differential Geometry'', v. 1-2, New-York - London, 1963.

\bibitem{gott}  J.R. Gott, {\it Phys. Rev. Lett.},  {\bf 66}, 1126 (1990).

\bibitem{morris}  M.S.Morris, K.S. Thorne and U. Yourtsever, {\it Phys. 
Rev.  Lett.} {\bf 61}, 1446 (1988).

\bibitem{idn}  I.D. Novikov, {\it ZhETF} {\bf 95}, 769 (1989).

\bibitem{frolnov}  V.P. Frolov and I.D. Novikov, {\it Phys. Rev. D.}
{\bf 42}, 1057 (1990).

\bibitem{visser}  M. Visser, {\it Phys. Rev. D}, {\bf 55}, 5212 (1997).

\bibitem{york}  J.W. York - in: ``Sources of Gravitational Radiation'', 
ed.  L.  Smarr, Cambridge, Cambridge U. P., 1979, p. 83.

\bibitem{fisher}  
A.E. Fisher and J.E. Marsden, ``Initial value
problem and dynamical formulation of general relativity'', in: ``General
Relativity. An Einstein Centenary Survey'', Eds. S.W. Hawking and W. 
Israel, Cambridge U.P., 1979.

\bibitem{myuk92}  M.Yu. Konstantinov, {\it Izvestiya Vuzov. Fizika}, 
1992, No. 12, 83.

\bibitem{myuk95}  M.Yu. Konstantinov, {\it Int. J. Mod. Phys. D}, 
{\bf v. 4}, 247 (1995).

\bibitem{thorne}  M.S. Morris and K.S. Thorne, {\it Amer. J. Phys.\/}
{\bf 59}, 395 (1988).

\bibitem{diaz}  Pedro F. Gonz\'ales-D\'\i az, {\it Phys. Rev. D.}
  {\bf 54}, 6122 (1996).

\bibitem{krasnikov}  S.V. Krasnikov, {\it Phys. Rev. D.} {\bf 55},
N 6, 3427 (1997).

\bibitem{carlini}  
A. Carlini, V.P. Frolov, M.B. Mensky, I.D. Novikov and H.H. Soleng, {\it 
Int.  J. Mod. Phys. D.} {\bf 4}, 557 (1995). 

\end{thebibliography}
\end{document}